\documentstyle[12pt,epsf]{article}

\newcommand{\beq}{\begin{equation}}
\newcommand{\eeq}{\end{equation}}
\newcommand{\beqa}{\begin{eqnarray}}
\newcommand{\eeqa}{\end{eqnarray}}
\newcommand{\beqan}{\begin{eqnarray*}}
\newcommand{\eeqan}{\end{eqnarray*}}

\newcommand{\ra}{\rightarrow}

\newcommand{\ben}{\begin{enumerate}}
\newcommand{\een}{\end{enumerate}}
\newcommand{\bfl}{\begin{flushleft}}
\newcommand{\efl}{\end{flushleft}}
\newcommand{\ba}{\begin{array}}
\newcommand{\ea}{\end{array}}
\newcommand{\btab}{\begin{tabular}}
\newcommand{\etab}{\end{tabular}}
\newcommand{\bit}{\begin{itemize}}
\newcommand{\eit}{\end{itemize}}

\newcommand{\cA}{{\cal A}}

\newcommand{\vs}{\vspace}
\newcommand{\hs}{\hspace}

\newcommand{\prepr}[1] {\begin{flushright}  {\bf #1} \end{flushright} \vskip
1.cm}
\newcommand{\titul}[1] {\begin{center}{\Large {\bf #1 } } \end{center}
\vskip 0.8cm}

\newcommand{\autor}[1] {\begin{center}  {\bf \lineskip .3cm #1  }
                        \end{center} }

\newcommand{\lugar}[1] {\begin{center}  {\normalsize \bf \it #1   } \end{center}}
\newcounter{muni}

\begin{document}
\hbadness=10000
\pagenumbering{arabic}
\begin{titlepage}
\prepr{
Preprint KEK-TH-626 \\ 
\hs{20mm} hep-ph/9905XXX }
\titul{\bf
Factorization and
Decay Constants $f_{D_s^*}$ and $f_{D_s}$}
\autor{ Dae Sung Hwang$^{1}$
\footnote{Email: dshwang@kunja.sejong.ac.kr}
 and Yong-Yeon Keum$^{2}$
\footnote{
Email: ccthmail.kek.jp; Monbusho Fellow in Japan
} } 

\lugar{ $^{1}$ Department of Physics, 
Sejong University, Seoul 143-747, Korea }

\lugar{ $^{2}$ Theory Group, KEK, Tsukuba, Ibaraki 305-0801 Japan }

\thispagestyle{empty}
%%%%%%%%%%%%%%%%%%%%%%%%%%%%%%%%%%%%%%%%%%%%%%%%%%%%%%%%%%%%%%%%%%%%%%%%%%%%%
\vs{10mm}
\begin{abstract}
\noindent{
We calculate the decay constants of $D_s$ and $D_s^{*}$ with
$\bar{B}^0 \ra D^{+}\ell^{-}\nu$ and $\bar{B}^0 \ra D^{+}D_s^{-(*)}$ decays.
In our analysis we take the factorization method with considering
non-factorizable term contributions
and used two
different form factor behaviours (constant and monopole-type)
for $F_0(q^2)$. 
We also consider the QCD-penguin
and Electroweak-penguin contributions in hadronic decays within the NDR
renormalization scheme at NLO calculation.
We estimate the decay constant of the $D_s$ meson to be $233\pm49$ ${\rm MeV}$
for (pole/pole)-type form factor and $255\pm54$ ${\rm MeV}$
for (pole/constant)-type form factor. For $D_s^{*}$ meson, we get
$f_{D_s^{*}} = 346 \pm 82$ ${\rm MeV}$, and $f_{D_s^{*}}/f_{D_s} = 1.43 \pm 0.45$
for (pole/constant)-type form factor.
} 

\vs{10mm}
{\rm  PACS index : 12.15.-y, 13.20.-v, 13.25.Hw, 14.40.Nd}

Keywards : Factorization, Non-leptonic Decays, Decay Constant, Penguin Effects
\end{abstract}

\thispagestyle{empty}
\end{titlepage}
\pagebreak

\baselineskip 22pt

\noindent
{\bf \large 1. Introduction}

Measuring purely leptonic decays of heavy mesons provides the most clear way for 
the determination of weak decay constants of heavy mesons, which connect 
the measured quantities, such as the $B\bar{B}$ mixing ratio, to CKM matrix
elements $V_{cb}, V_{ub}$.
However, currently it is not possible to determine $f_B, f_{B_s}$ 
$f_{D_s}$
and
$f_{D^{*}_s}$
experimentally from leptonic $B$ and $D_s$ decays.
For instance, the decay rate for $D_s^{+}$ is given by \cite{rosner}
\beq
\Gamma(D_s^{+} \ra \ell^{+} \nu) = {G_F^2 \over 8 \pi} f_{D_s}^2 m_{\ell}^2 M_{D_s}
\left( 1 - {m_{l}^2 \over M_{D_s}^2} \right)^2 |V_{cs}|^2
\eeq
Because of helicity suppression, the electron mode $D_s^{+} \ra e^{+}\nu$
has a very small rate. The relative widths are $10 : 1 : 2 \times 10^{-5}$ for
$\tau^{+}\nu, \mu^{+}\nu$ and $e^{+}\nu$ final states, respectively.
Unfortunately the mode with the largest branching fraction, $\tau^{+}\nu$,
has at least two neutrinos in the final state and is difficult to detect
in experiment. 
So theoretical calculations for decay constant have to be used.
The factorization ansatz for nonleptonic decay modes provides us a good
approximate method to obtain nonperturbative quantities such as form factors
and decay constants which are hardly accessible in any other way
\cite{bort-stone,HF}.

There are many ways that the quarks produced in a nonleptonic weak decay can arrange
themselves into hadrons. 
The final state is linked to the initial state by complicated trees of gluon and 
quark interactions, pair production, and loops. These make the theoretical 
description of nonleptonic decays difficult.
However, since the products of a $B$ meson decay are quite energetic,
it is possible that the complicated QCD interactions are less important and 
that the two quark pairs of the currents in the weak Hamiltonian
group individually into the final state mesons without further exchanges of gluons.
The color transparency argument suggests that a quark-antiquark pair remains 
at a state of small size with a correspondingly small chromomagnetic moment 
until it is far from the other decay products.

Color transparency is the basis for the factorization hypothesis, in which
amplitudes factorize into products of two current matrix elements.
This ansatz is widely used in heavy quark physics, 
as it is almost the only way to treat hadronic decays.

In this paper we consider the way how to determine weak decay constants
$f_{D_s}$ and $f_{D_s^{*}}$ under factorization ansatz including
penguin effects.
In section 2 we discuss the way how to extract the unknown parameter
$|V_{cb}F_1^{BD}(0)|$ from the branching ratio of the semileptonic decay
$\bar{B}^{0} \ra D^{+}\ell \bar{\nu}$.
In order to check the validity of the factorization assumption,
we study the nonleptonic two body decays, 
$B \ra D\rho, D\pi$ and $DK^{(*)}$ in section 3.
In section 4 we calculate $f_{D_s}$
and $f_{D_s^{*}}$ from $\bar{B}^0 \ra D^{+}D_s^{-(*)}$ decay modes.
In our analysis we improve the previous analysis \cite{YYKeum} 
by considering  the QCD-penguin and Electroweak-penguin effects of
about 13 $\%$ for $B \ra D D_s$ and 4 $\%$ for $B \ra D D_s^{*}$,
which are not negligible as  discusssed in \cite{GKKP}.
Also we follow the gauge independent approach to calculate the
effective Wilson coefficients which was studied by 
pertubative QCD factorization theorem \cite{Hai-Yang}.

\vs{15mm}
{\bf \large 2. Semileptonic Decay
${\bar{B}}^0\rightarrow D^+ l^- {\bar{\nu}}$} 

{}From Lorentz invariance one finds the decomposition of the
hadronic matrix element in terms of hadronic form factors:
\begin{eqnarray}
<D^+(p_D )|J_\mu |{\bar{B}}^0(p_B)>
&=&
\left[ (p_B+p_D )_\mu 
-{m_B^2-m_D^2\over q^2}q_\mu \right] \, F_1^{BD}(q^2)
\nonumber \\
\cr
&&
\hs{3mm}
+{m_B^2-m_D^2\over q^2}\, q_\mu \, F_0^{BD}(q^2),
\label{a1}
\end{eqnarray}
where $J_\mu = {\bar{c}}\gamma_\mu b$ and
$q_\mu =(p_B-p_D )_\mu$.
In the rest frame of the decay products, $F_1(q^2)$ and $F_0(q^2)$
correspond to $1^-$ and $0^+$ exchanges, respectively.
At $q^2=0$ we have the constraint
\begin{equation}
F_1^{BD}(0)=F_0^{BD}(0),
\label{a1a}
\end{equation}
since the hadronic matrix element in (\ref{a1}) is nonsingular
at this kinematic point.

The $q^2$ distribution in the semileptonic decay
${\bar{B}}^0\rightarrow D^+ l^- {\bar{\nu}}$
is written in terms of the hadronic form factor
$F_1^{BD}(q^2)$ as
\begin{equation}
{d\Gamma ({\bar{B}}^0\rightarrow D^+ l^- {\bar{\nu}})\over dq^2}
={G_F^2\over 24\pi^3}\, |V_{cb}|^2\, [K(q^2)]^3\,
|F_1^{BD}(q^2)|^2,
\label{a2}
\end{equation}
where the $q^2$ dependent momentum $K(q^2)$ is given by
\begin{equation}
K(q^2)={1\over 2m_B}\,
\left[ (m_B^2+m_D^2-q^2)^2
-4m_B^2m_D^2 \right]^{1/2}.
\label{a3}
\end{equation}
In the zero lepton mass limit,
$0\le q^2\le (m_B-m_D)^2$.

For the $q^2$ dependence of the form factors,
Wirbel et al. \cite{wsb} assumed a simple pole formula
for both $F_1(q^2)$ and $F_0(q^2)$ (pole/pole):
\begin{equation}
F_1(q^2)=F_1(0)\,/ ( 1-{q^2\over m_{F_1}^2}),\qquad
F_0(q^2)=F_0(0)\,/ ( 1-{q^2\over m_{F_0}^2}),
\label{a4}
\end{equation}
with the pole masses
\begin{equation}
m_{F_1}=6.34\ {\rm GeV},\qquad
m_{F_0}=6.80\ {\rm GeV}.
\label{a4aa}
\end{equation}
Korner and Schuler \cite{ks} also adopted the same $q^2$ dependence of
$F_1(q^2)$ and $F_0(q^2)$ given by (\ref{a4}) and (\ref{a4aa}).
On the other hand, the heavy quark effective theory (HQET)
gives in the $m_{b,c}\rightarrow \infty$ limit
the relation between $F_1(q^2)$ and $F_0(q^2)$
given by \cite{iw,nr}
\begin{equation}
F_0(q^2)=\left[ 1-{q^2\over (m_B+m_D)^2}\right] \hs{3mm} F_1(q^2).
\label{a4a1}
\end{equation}
The combination of (\ref{a4}) and (\ref{a4a1}) suggests that $F_0(q^2)$
is approximately constant when we keep the simple pole dependence for
$F_1(q^2)$.
Therefore, in this paper, as well as the above (pole/pole) form factors,
we will also consider the following ones (pole/const.):
\begin{equation}
F_1(q^2)=F_1(0)\,/( 1-{q^2\over m_{F_1}^2}),\qquad
F_0(q^2)=F_0(0),
\label{a4b1}
\end{equation}
with
\begin{equation}
m_{F_1}=6.34\ {\rm GeV}.
\label{a4aab2}
\end{equation}

By introducing the variable
$x\equiv q^2/m_B^2$,
which has the range of
$0\le x\le (1-{m_D\over m_B})^2$
in the zero lepton mass limit,
(\ref{a2}) is written as
\begin{eqnarray}
{d\Gamma ({\bar{B}}^0\rightarrow D^+ l^- {\bar{\nu}})\over dx}
&=&{G_F^2m_B^5\over 192\pi^3}\, |V_{cb}\, F_1^{BD}(0)|^2\,
{{\lambda}^{3}[1,\, {m_D^2\over m_B^2},\, x] \over
\Bigl( 1-{m_B^2\over m_{F_1}^2}x{\Bigr)}^2},
\label{a5}\\
{\lambda}[1,\, {m_D^2\over m_B^2},\, x]&=&
\left[ (1+{m_D^2\over m_B^2}-x)^2-4{m_D^2\over m_B^2}
\right]^{1/2}.
\nonumber
\end{eqnarray}
Then the branching ratio
${\cal{B}} ({\bar{B}}^0\rightarrow D^+ l^- {\bar{\nu}})$
is given by
\begin{eqnarray}
{\cal{B}} ({\bar{B}}^0\rightarrow D^+ l^- {\bar{\nu}})&=&
({G_Fm_B^2\over {\sqrt{2}}})^2\, {m_B\over {\Gamma}_B}\,
{2\over 192 {\pi}^2}\,
|V_{cb}\, F_1^{BD}(0)|^2\times I
\nonumber\\
&=&
2.221\times 10^2 \,\, |V_{cb}\, F_1^{BD}(0)|^2\times I,
\label{a6}
\end{eqnarray}
where the dimensionless integral $I$ is given by
\begin{equation}
I=
\int_0^{(1-{m_D\over m_B})^2} dx
{ \left[ (1+{m_D^2\over m_B^2}-x)^2
-4{m_D^2\over m_B^2} \right]^{3/2}\over
\Bigl( 1-{m_B^2\over m_{F_1}^2}x{\Bigr)}^2}
=0.121
\label{a7}
\end{equation}
In obtaining the numerical values in (\ref{a6}) and (\ref{a7}),
we used the following
experimental results \cite{rpp}:
$m_D=m_{D^+}=1.869$ GeV, $m_B=m_{B^0}=5.279$ GeV,
${\Gamma}_B={\Gamma_{B^0}}=4.219\times 10^{-13}$ GeV
(${\tau}_{B^0}=(1.56\pm 0.06)\times 10^{-12}$ s),
and $G_F=1.166\ 39(2)\times 10^{-5}\ {\rm GeV}^{-2}$.
Since
${\cal{B}} ({\bar{B}}^0\rightarrow D^+ l^- {\bar{\nu}})=
(1.78\pm 0.20\pm 0.24)\times 10^{-2}$ was obtained experimentally,
the value of $|V_{cb}\, F_1^{BD}(0)|$ can be extracted from (\ref{a6}).
Following this procedure, we obtain \cite{stone}
\begin{equation}
|V_{cb}\, F_1^{BD}(0)|=(2.57\pm 0.14\pm 0.17)\times 10^{-2}.
\label{a7a}
\end{equation}
In the calculations of the next sections,
we will use $|V_{cb}\, F_1^{BD}(0)|=(2.57\pm 0.22)\times 10^{-2}$
which is given by combining the statistical and systematic errors in
(\ref{a7a}).

\vs{15mm}
{\bf \large 3. Test of Factorization with
${\bar{B}}^0\rightarrow D^+ \rho^-$ and
${\bar{B}}^0\rightarrow D^+ \pi^-$,
and Prediction of Branching Ratio
${\cal{B}}({\bar{B}}^0\rightarrow D^+ K^{-(*)})$
} \\

In general the test of factorization, independent of the numerical values
of $a_1$, $a_2$ and of the CKM parameters $|V_{cb}|$ or $|V_{ub}|$,
can be carried out by considering the ratios of rates for two Class I
or Class II $B$-meson hadronic two-body decays.
On the other hand, we can also use the relation between the
semi-leptonic decays and the non-leptonic decays with $a_1$ and $a_2$
given by other sources.
In our analysis we use the latter one.

Let us start by recalling the relevant effective weak Hamiltonian:
\begin{equation}
{\cal H}_{\rm eff}={G_F\over {\sqrt{2}}}V_{cb}V_{ud}^*
[C_1(\mu ){\cal O}_1+C_2(\mu ){\cal O}_2]\ +\ {\rm H.C.},
\label{b1}
\end{equation}
\begin{equation}
{\cal O}_1=({\bar{d}}\Gamma^\rho u)({\bar{c}}\Gamma_\rho b),\quad
{\cal O}_2=({\bar{c}}\Gamma^\rho u)({\bar{d}}\Gamma_\rho b),
\label{b2}
\end{equation}
where $G_F$ is the Fermi coupling constant, $V_{cb}$ and $V_{ud}$
are corresponding Cabibbo-Kobayashi-Maskawa (CKM) matrix elements
and $\Gamma_\rho = \gamma_\rho (1-\gamma_5)$.
The Wilson coefficients $C_1(\mu )$ and $C_2(\mu )$ incorporate
the short-distance effects arising from the renormalization of
${\cal H}_{\rm eff}$ from $\mu =m_W$ to $\mu =O(m_b)$.
By using the Fierz transformation under which $V-A$ currents remain
$V-A$ currents, we get the following equivalent forms:
\begin{eqnarray}
C_1{\cal O}_1+C_2{\cal O}_2&=&
(C_1+{1\over N_c}C_2){\cal O}_1
+C_2({\bar{d}}\Gamma^\rho T^au)({\bar{c}}\Gamma_\rho T^ab)
\nonumber\\
&=&(C_2+{1\over N_c}C_1){\cal O}_2
+C_1({\bar{c}}\Gamma^\rho T^au)({\bar{d}}\Gamma_\rho T^ab),
\label{b3}
\end{eqnarray}
where $N_c=3$ is the number of colors and $T^a$'s are $SU(3)$ color
generators.
The second terms in (\ref{b3}) involve color-octet currents.
In the factorization assumption, these terms are neglected and
${\cal H}_{\rm eff}$ is rewritten in terms of ``factorized hadron
operators'' \cite{wsb}:
\begin{equation}
{\cal H}_{\rm eff}={G_F\over {\sqrt{2}}}V_{cb}V_{ud}^*
\Bigl( a_1[{\bar{d}}\Gamma^\rho u]_H[{\bar{c}}\Gamma_\rho b]_H
+a_2[{\bar{c}}\Gamma^\rho u]_H[{\bar{d}}\Gamma_\rho b]_H\Bigr)
\ +\ {\rm H.C.},
\label{b4}
\end{equation}
where the subscript $H$ stands for $hadronic$ implying that the
Dirac bilinears inside the brackets be treated as interpolating
fields for the mesons and no further Fierz-reordering need be done.
The phenomenological parameters $a_1$ and $a_2$ are related to
$C_1$ and $C_2$ by
\begin{equation}
a_1=C_1+{1\over N_c}C_2,\quad
a_2=C_2+{1\over N_c}C_1.
\label{b5}
\end{equation}
{}From the analyses of A.J. Buras \cite{burasNDR}, the parameters $a_1$ and
$a_2$ are determined at NLO calculation in the NDR scheme as
\begin{equation}
a_1=1.02\pm 0.01,\quad a_2=0.20\pm 0.05.
\label{b6}
\end{equation}

For the two body decay, in the rest frame of initial meson
the differential decay rate is given by
\begin{equation}
d\Gamma ={1\over 32\pi^2}|{\cal M}|^2
{|{\bf p}_1|\over M^2}d\Omega ,
\label{b7}
\end{equation}
\begin{equation}
|{\bf p}_1|=
{[(M^2-(m_1+m_2)^2)(M^2-(m_1-m_2)^2)]^{{1/2}}\over 2M},
\label{b8}
\end{equation}
where $M$ is the mass of initial meson, and $m_1$ ($m_2$) and
${\bf p}_1$ are the mass and momentum of one of final mesons.
By using (\ref{a1}), (\ref{b4}) and
$<0|\Gamma_\mu |\rho(q,\varepsilon )>
=\varepsilon_\mu (q) m_{\rho} f_{\rho}$,
(\ref{b7}) gives the following formula for the branching ratio of
${\bar{B}}^0\rightarrow D^+ \rho^-$:
\begin{eqnarray}
{\cal{B}} ({\bar{B}}^0\rightarrow D^+ \rho^-)
&=& \left({G_Fm_B^2\over {\sqrt{2}}}\right)^2\,
|V_{ud}|^2\, {1\over 16 \pi }\, {m_B\over {\Gamma}_B}\,
a_1^2\,
{f_{\rho}^2 \over m_B^2}\, |V_{cb}\, F_1^{BD}(m_{\rho}^2)|^2
\nonumber\\
&\times &
\left[\Bigl( 1-({m_D+m_{\rho}\over m_B})^2{\Bigr)}
 \Bigl( 1-({m_D-m_{\rho}\over m_B})^2{\Bigr)}
\right]^{3/2}
\nonumber\\
&=&13.25
\times |V_{cb}\, F_1^{BD}(m_{\rho}^2)|^2
\times \left( {a_1 \over 1.02} \right)^2.
\label{b12}
\end{eqnarray}
In obtaining the numerical values in (\ref{b12}), we used the
experimental results given below (\ref{a7}),
$m_{\rho}=m_{\rho^+}=766.9$ MeV,
$f_{\rho}=f_{\rho^+}=216$ MeV,
and $V_{ud}=0.9751$ \cite{rpp}.
For the value of $a_1$ we used the value given in (\ref{b6}).
Then, by using the formula (\ref{b12}) with the values
of $|V_{cb}\, F_0^{BD}(0)|^2$ $(F_0^{BD}(0)=F_1^{BD}(0))$
given in (\ref{a7a}),
we obtain the branching ratio
${\cal{B}}({\bar{B}}^0\rightarrow D^+ \rho^-)$ presented in Table 
\ref{table1}.

For the process ${\bar{B}}^0\rightarrow D^+ K^{*-}$,
by using
$<0|\Gamma_\mu |K^{*}(q,\varepsilon )>
=\varepsilon_\mu (q) m_{K^{*}} f_{K^{*}}$,
we have
\begin{eqnarray}
{\cal{B}} ({\bar{B}}^0\rightarrow D^+ K^{*-})
&=& \left({G_Fm_B^2\over {\sqrt{2}}}\right)^2\,
|V_{us}|^2\, {1\over 16\pi }\, {m_B\over {\Gamma}_B}\,
a_1^2\,
{f_{K^*}^2 \over m_B^2}\, |V_{cb}\, F_1^{BD}(m_{K^*}^2)|^2
\nonumber\\
&\times &
\left[\Bigl( 1-({m_D+m_{K^*}\over m_B})^2{\Bigr)}
 \Bigl( 1-({m_D-m_{K^*}\over m_B})^2{\Bigr)}
\right]^{3/2}
\nonumber\\
&=&0.67\times
|V_{cb}\, F_1^{BD}(m_{K^*}^2)|^2
\times \left( {a_1 \over 1.02} \right)^2.
\label{b12a}
\end{eqnarray}
where we used $m_{K^*}=m_{K^{*-}}=891.59$ MeV,
$f_{K^*}=f_{K^{*-}}=218$ MeV,
and $V_{us}=0.2215$ \cite{rpp}.
By using (\ref{b12a}) with $|V_{cb}\, F_1^{BD}(0)|^2$ in (\ref{a7a}),
we obtain the branching ratio
${\cal{B}}({\bar{B}}^0\rightarrow D^+ K^{*-})$ presented in Table \ref{table1}.

By using (\ref{a1}), (\ref{b4}) and
$<0|\Gamma_\mu |\pi(q)>=iq_\mu f_{\pi}$,
(\ref{b7}) gives the following formula for the branching ratio of the
process ${\bar{B}}^0\rightarrow D^+ \pi^-$:
\begin{eqnarray}
{\cal{B}} ({\bar{B}}^0\rightarrow D^+ \pi^-)
&=&\left({G_Fm_B^2\over {\sqrt{2}}}\right)^2\,
|V_{ud}|^2\, {1\over 16 \pi }\, {m_B\over {\Gamma}_B}\, a_1^2\,
{f_{\pi}^2 \over m_B^2}\, |V_{cb}\, F_0^{BD}(m_{\pi}^2)|^2
\nonumber\\
& &\times \Bigl( 1-{m_D^2\over m_B^2}{\Bigr)}^2\,
\left[\Bigl( 1-({m_D+m_{\pi}\over m_B})^2{\Bigr)}
\Bigl( 1-({m_D-m_{\pi}\over m_B})^2{\Bigr)}\right]^{1/2}
\nonumber\\
&=&5.42\times
|V_{cb}\, F_0^{BD}(m_{\pi}^2)|^2
\times \left( {a_1 \over 1.02} \right)^2,
\label{b11}
\end{eqnarray}
where we used
$m_{\pi}=m_{\pi^-}=139.57$ MeV and
$f_{\pi}=f_{\pi^-}=131.74$ MeV \cite{rpp}.
By using the formula (\ref{b11}) with the values
of $|V_{cb}\, F_0^{BD}(0)|^2$ $(F_0^{BD}(0)=F_1^{BD}(0))$
in (\ref{a7a}),
we obtain the branching ratio
${\bar{B}}^0\rightarrow D^+ \pi^-$ presented in Table \ref{table1}.

For the process ${\bar{B}}^0\rightarrow D^+ K^-$,
by using $<0|\Gamma_\mu |K^-(q)>=iq_\mu f_{K^-}$,
we have
\begin{eqnarray}
{\cal{B}} ({\bar{B}}^0\rightarrow D^+ K^-)
&=& \left({G_Fm_B^2\over {\sqrt{2}}}\right)^2\,
|V_{us}|^2\, {1\over 16\pi }\, {m_B\over {\Gamma}_B}\, a_1^2\,
{f_{K}^2 \over m_B^2}\, |V_{cb}\, F_0^{BD}(m_{K}^2)|^2
\nonumber\\
& &\times \Bigl( 1-{m_D^2\over m_B^2}{\Bigr)}^2\,
\left[\Bigl( 1-({m_D+m_{K}\over m_B})^2{\Bigr)}
\Bigl( 1-({m_D-m_{K}\over m_B})^2{\Bigr)}
\right]^{1/2}
\nonumber\\
&=&0.41\times
|V_{cb}\, F_0^{BD}(m_{K}^2)|^2
\times \left( {a_1 \over 1.02} \right)^2.
\label{b11a}
\end{eqnarray}
where we used $m_{K}=m_{K^-}=493.68$ MeV,
$f_{K}=f_{K^+}=160.6 {\rm MeV}$ \cite{rpp}.
By using (\ref{b11a}) with $|V_{cb}\, F_1^{BD}(0)|^2$ in (\ref{a7a}),
we obtain the branching ratio
${\cal{B}}({\bar{B}}^0\rightarrow D^+ K^-)$ presented in Table \ref{table1}.
It seems that the factorization method works well in $\bar{B}^0 \rightarrow
D^{+}\pi^{-}, D^{+}\rho^{-}$ decays. We predict branching ratios :
\begin{eqnarray} 
&& {\cal B}(\bar{B}^0 \rightarrow D^{+}K^{-}) \simeq 2.7 \cdot 10^{-4} 
\cdot \left({a_1 \over 1.02}\right)^2
\nonumber \\
&& {\cal B}(\bar{B}^0 \rightarrow D^{+}K^{*-}) \simeq 4.6 \cdot
10^{-4}
\cdot \left({a_1 \over 1.02}\right)^2
\end{eqnarray}
which is certainly reachable in near future.

\vs{15mm}
{\bf \large 4. Determination of $f_{D_s^{*}}$
and $f_{D_s}$ from
${\bar{B}}^0\rightarrow D^+ D_s^{-*}$ and
${\bar{B}}^0\rightarrow D^+ D_s^-$} \\

The effective Hamiltonian for $\triangle B = 1$ transitions is given by
\begin{equation}
H_{\rm eff}=
{G_F\over {\sqrt{2}}}[V_{ub}V_{uq}^*(C_1O_1^u+C_2O_2^u)
+V_{cb}V_{cq}^*(C_1O_1^c+C_2O_2^c)
-V_{tb}V_{tq}^*\sum_{i=3}^6C_iO_i],
\label{pb1}
\end{equation}
where $q=d,s$ and $C_i$ are the Wilson coefficients evaluated at the
renormalization scale $\mu$, and
the current-current operators $O_1^{u,c}$ and $O_2^{u,c}$ are
\begin{eqnarray}
O_1^u=({\bar{u}}_\alpha b_\alpha )_{V-A}({\bar{q}}_\beta u_\beta )_{V-A}
&\qquad &
O_1^c=({\bar{c}}_\alpha b_\alpha )_{V-A}({\bar{q}}_\beta c_\beta )_{V-A}
\nonumber\\
O_2^u=({\bar{u}}_\beta b_\alpha )_{V-A}({\bar{q}}_\alpha u_\beta )_{V-A}
&\qquad &
O_2^c=({\bar{c}}_\beta b_\alpha )_{V-A}({\bar{q}}_\alpha c_\beta )_{V-A},
\label{pb2}
\end{eqnarray}
and the QCD penguin operators $O_3\ -\ O_6$ are
\begin{eqnarray}
O_3=({\bar{q}}_\alpha b_\alpha )_{V-A}
\sum_{q'}({\bar{q}}'_\beta q'_\beta )_{V-A}
&\qquad &
O_4=({\bar{q}}_\beta b_\alpha )_{V-A}
\sum_{q'}({\bar{q}}'_\alpha q'_\beta )_{V-A}
\nonumber\\
O_5=({\bar{q}}_\alpha b_\alpha )_{V-A}
\sum_{q'}({\bar{q}}'_\beta q'_\beta )_{V+A}
&\qquad &
O_6=({\bar{q}}_\beta b_\alpha )_{V-A}
\sum_{q'}({\bar{q}}'_\alpha q'_\beta )_{V+A}.
\end{eqnarray}
The electroweak penguin operators $O_7\ -\ O_{10}$ are given by :
\begin{eqnarray}
O_7=({\bar{q}}_\alpha b_\alpha )_{V-A}
\sum_{q'} {3 \over 2} e_{q'}
({\bar{q}}'_\beta q'_\beta )_{V+A}
&\qquad &
O_8=({\bar{q}}_\beta b_\alpha )_{V-A}
\sum_{q'} {3 \over 2} e_{q'}
({\bar{q}}'_\alpha q'_\beta )_{V+A}
\nonumber \\
O_9=({\bar{q}}_\alpha b_\alpha )_{V-A}
\sum_{q'} {3 \over 2} e_{q'}
({\bar{q}}'_\beta q'_\beta )_{V-A}
&\qquad &
O_8=({\bar{q}}_\beta b_\alpha )_{V-A}
\sum_{q'} {3 \over 2} e_{q'}
({\bar{q}}'_\alpha q'_\beta )_{V-A}.
\label{pb3}
\end{eqnarray}
In (\ref{pb1}) we consider the effects of the electroweak penguin operators,
however, we neglect the contribution of the dipole operators, 
since its contribution is not important in this work.

When we take $m_t=174$ GeV, $m_b=5.0$ GeV, $\alpha_{\rm s}(M_z)=0.118$
and $\alpha_{\rm em}(M_z)=1/128$, the numerical values of the renormalization
scheme independent Wilson coefficients ${\bar{C}}_i$ at $\mu =m_b$ are
given by \cite{deshpande}
\begin{eqnarray}
&&{\bar{C}}_1=-0.3125,\ \ \ {\bar{C}}_2=1.1502,
\nonumber\\
&&{\bar{C}}_3=0.0174,\ \ \ {\bar{C}}_4=-0.0373,\ \ \
{\bar{C}}_5=0.0104,\ \ \ {\bar{C}}_6=-0.0459,
\nonumber \\
&&{\bar{C}}_7=-1.050 \times 10^{-5},\ \ \ 
{\bar{C}}_8=3.839 \times 10^{-4}, 
\nonumber \\
&&{\bar{C}}_9=-0.0101,\ \ \ 
{\bar{C}}_{10}=1.959 \times 10^{-3}.
\label{pb4}
\end{eqnarray}
The effective Hamiltonian in (\ref{pb1}) for the decays
${\bar{B}}^0\rightarrow D^+ D_s^{-(*)}$
can be rewritten as
\begin{equation}
H_{eff}=
{G_F\over {\sqrt{2}}}
[V_{cb}V_{cs}^*(C_1^{eff}O_1^c+C_2^{eff}O_2^c)
-V_{tb}V_{ts}^*\sum_{i=3}^{10}C_i^{eff}O_i],
\label{pb5}
\end{equation}
where $C_i^{eff}$ are given by \cite{fleischer}
\begin{eqnarray}
&&C_1^{eff}={\bar{C}}_1,\ \ \ C_2^{eff}={\bar{C}}_2,\ \ \
C_3^{eff}={\bar{C}}_3-P_{\rm s}/N_{\rm c},\ \ \
C_4^{eff}={\bar{C}}_4+P_{\rm s},
\nonumber\\
&&C_5^{eff}={\bar{C}}_5-P_{\rm s}/N_{\rm c},\ \ \
C_6^{eff}={\bar{C}}_6+P_{\rm s},\ \ \
C_7^{eff}={\bar{C}}_7+P_{\rm e},\ \ \
C_8^{eff}={\bar{C}}_8,
\nonumber\\
&&C_9^{eff}={\bar{C}}_9+P_{\rm e},\ \ \
C_{10}^{eff}={\bar{C}}_{10}.
\label{pb6}
\end{eqnarray}
with
\begin{eqnarray}
&&P_{\rm s}={\alpha_{\rm s}\over 8\pi }
[{10\over 9}-G(m_q,q^2,\mu )]{\bar{C}}_2(\mu ),
\nonumber\\
&&P_{\rm e}={\alpha_{\rm em} \over 9\pi }
[{10\over 9}-G(m_q,q^2,\mu )]
(3 \bar{C}_{1}(\mu) + \bar{C}_2(\mu )),
\label{pb7}
\\
&&G(m_q,q^2,\mu )=-4\int_0^1x(1-x)\, 
{\rm ln}({m_q^2-x(1-x)q^2\over \mu^2})dx,
\nonumber \\
\end{eqnarray}
where $q$ denotes the momentum of the virtual gluons appearing in the QCD 
time-like matrix elements, and $N_{\rm c}$ is the number of colors.
Assuming $q^2 = m_b^2/2$, we obtain the analytic formular for $G(m_q,q^2,\mu)$ :
\beq
G(m_q,{m_b^2 \over 2},\mu = m_b) 
= -{2 \over 3} {\rm ln}\left({y \over 8} \right) + {10 \over 9}
+{2 \over 3} y + {(2 + y)\sqrt{1 - y} \over 3}
\left[ {\rm ln}\left|{1 - \sqrt{1 - y} \over 1 
+ \sqrt{1 - y}} \right| + i \pi \right]
\label{pb8}
\eeq
with $y = 8 m_q^2/m_b^2$. 

By considering the non-factorizable term contributions,
the relation between the effective coefficients $a_{i}^{eff}$ and
the Wilson coefficients in the effective Hamiltonian are given by
\begin{equation}
a^{eff}_{2i} = C_{2i}^{eff} + {1 \over N_c^{eff}} C_{2i -1}^{eff},\\\
\hspace{10mm}
a^{eff}_{2i -1} = C_{2i -1}^{eff} + {1 \over N_c^{eff}} C_{2i}^{eff}.
\end{equation}
where $i = 1,...,5$, and the non-factorizable effects are absorbed
into the $N_c^{eff}$ by
\begin{equation}
{1 \over N_c^{eff}}_{i} \equiv {1 \over N_c} + \chi_{i}, \hs{10mm}
N_c = 3.
\end{equation}
In order to simplify the notation, we will use the notation $a^{i}$
instead of $a^{eff}_{i}$ in the below.

In usual factorization approach, when we consider the off-shell momentum
of the external quark line,
the effective Wilson coefficients has the
ambiguities of the infrared cutoff and gauge dependence.
As stressed by \cite{BS}, the gauge and infrared dependence always
appears as long as the matrix elements of operators are calculated 
between quark states.
Recently this problem was sloved by pertubative QCD factorization theorm
\cite{Hai-Yang} by using the on-shell external quark.
By  following their approach and
inserting the values for $m_q = m_c(\mu) = 0.95$ ${\rm GeV}$,
we get the values $C_{i}^{eff} (i = 1 \sim 10)$ for $b \rightarrow c$
given in Table \ref{table2}.
For different combinations of $N_c^{eff} = 2, 3$, and $5$,
the values of the effective coefficients $a_{i} (i = 1 \sim 10)$ 
are shown in 
Table \ref{table3}.
Here $(N_c)_{LL,LR} = 3$ corresponds to the naive factorization
 approximation without considering non-factorizable contributions.

The decay amplitude
${\cal A}({\bar{B}}^0\rightarrow D^+ D_s^-)
\equiv <D^+ D_s^-|{\cal{H}}_{\rm eff}|{\bar{B}}^0>$
is given as follows:
\beqa
{\cal A}({\bar{B}}^0\rightarrow D^+ D_s^-)
&=& {G_F\over {\sqrt{2}}}
[V_{cb}V_{cs}^*a_1 - V_{tb}V_{ts}^*( a_4 +a_{10}
+2 (a_6 + a_8) {m_{D_s}^2\over (m_b-m_c)(m_s+m_c)})] \hs{2mm} {\cal M}_{a}
\nonumber \\
\cr
& \simeq &
{G_F\over {\sqrt{2}}} V_{cb}V_{cs}^* \,\, a_1
\,\, R_{DDs} \,\, {\cal M}_{a}
\label{pb9}
\eeqa
where
\beq
R_{DDs} =
\left[1 + {(a_4 + a_{10}) \over a_1} + 2 {(a_6 + a_8) \over a_1}
{m_{D_s}^2 \over (m_b-m_c)(m_s+m_c)} \right]  
\label{pb9a}
\eeq
and
\beq
{\cal M}_a =  <D_s^-|{\bar{s}}\gamma^\mu\gamma_5c|0>
<D^{+}|{\bar{c}}\gamma_\mu b|{\bar{B}}^0>
= - i f_{D_s}(m_B^2 - m_D^2) F_0^{BD}(m_{D_s}^2)
\label{pb9b}
\eeq 

On the other hand, we have
\beqa
{\cal A}({\bar{B}}^0\rightarrow D^+ D_s^{-*})
&=& {G_F\over {\sqrt{2}}}
[V_{cb}V_{cs}^*a_1 -V_{tb}V_{ts}^* (a_4 + a_{10})] \hs{2mm}
{\cal M}_b \nonumber \\
\cr
& \simeq  &  {G_F\over {\sqrt{2}}}
 V_{cb}V_{cs}^* \,\, a_1 \,\,
R_{DDs^{*}} \hs{2mm} {\cal M}_b
\label{pb10}
\eeqa
where
\beq
R_{DDs^{*}} =
\left(1 + {(a_4 + a_{10}) \over a_1} \right) 
\label{pb10a}
\eeq
and 
\beq
{\cal M}_b =  <D_s^{*}|{\bar{s}}\gamma^\mu\gamma_5c|0>
<D^{+}|{\bar{c}}\gamma_\mu b|{\bar{B}}^0>
= m_{D_s^{*}}f_{D_s^{*}}[\epsilon(q) \cdot (p_B + p_D)] F_1^{BD}(m_{D_s^{*}}^2).
\eeq
 
We can estimate the penguin contributions for each process, for exapmle,
in the case of $N_{LL}=2$ and $N_{LR} = 5$ :
\beqa
{\rm For} \hs{1mm} \bar{B}^0 \ra D^{+} D_s^{-} ; & &
\left| {\cA_P \over \cA_T} \right| =
\left|{(a_4 + a_{10}) \over a_1} + 2 {(a_6 + a_8) \over a_1}
{m_{D_s}^2 \over (m_b - m_c) (m_c + m_s)} \right| = 13.1 \% \\
\cr
{\rm For} \hs{2mm} \bar{B}^0 \ra D^{+}D_s^{*-} ; &&
\left|{\cA_P \over \cA_T} \right| =
\left|{(a_4 + a_{10}) \over a_1} \right| =  3.9 \%
\eeqa
where $\cA_T(\cA_P )$ stands for the amplitude of tree diagram 
(penguin diagram).
Here we used the values $m_c(m_b) = 0.95$ ${\rm GeV}$ and
$m_s(m_b) = 90$ ${\rm MeV}$.
Therefore, the penguin contributions affect the extraction of the decay constants
$f_{D_s}$ and $f_{D^{*}_s}$.
The penguin contributions for $B \ra DD_s$ is more than three times of
those for $B \ra DD_s^{*}$.

{}From (\ref{pb9}) and (\ref{pb10})
the decay constants are given by
\begin{eqnarray}
f_{D_s^{*}}&=&(0.87\times 10^{-1}\ {\rm GeV})\cdot
{\sqrt{{\cal{B}}({\bar{B}}^0\rightarrow D^+ D_s^{-*})}
\over
|V_{cb}\, F_1^{BD}(m_{D_s^{*}}^2)|}
\cdot \left({1.02 \over a_1}\right) \cdot {1 \over R_{DDs^{*}}},
\nonumber \\
\cr
f_{D_s}&=&(0.64\times 10^{-1}\ {\rm GeV})\cdot
{\sqrt{{\cal{B}}({\bar{B}}^0\rightarrow D^+ D_s^{-})}
\over
|V_{cb}\, F_0^{BD}(m_{D_s}^2)|}
\cdot \left({1.02 \over a_1}\right) \cdot {1 \over R_{DDs}}.
\label{c11}
\end{eqnarray}
Browder et al. \cite{browder} presented the following experimental
results for the branching ratios:
\begin{eqnarray}
{\cal{B}}({\bar{B}}^0\rightarrow D^+ D_s^{-*})&=&
(1.14\pm 0.42\pm 0.28)\times 10^{-2}
=(1.14\pm 0.50)\times 10^{-2},
\nonumber\\
{\cal{B}}({\bar{B}}^0\rightarrow D^+ D_s^{-})&=&
(0.74\pm 0.22\pm 0.18)\times 10^{-2}
=(0.74\pm 0.28)\times 10^{-2},
\label{c12}
\end{eqnarray}
where we combined the statistical and systematic errors.
{}From (\ref{pb9a}),(\ref{pb10a}),(\ref{c11}), and (\ref{c12}), 
we obtain the results which are obtained by including the penguin contributions:
\begin{eqnarray}
f_{D_s^*}=346\pm 82\ {\rm MeV},\qquad f_{D_s}=233\pm 49\ {\rm MeV}& &
{\rm for\ (pole/pole)},
\nonumber\\
f_{D_s^*}=346\pm 82\ {\rm MeV},\qquad f_{D_s}=255\pm 54\ {\rm MeV}& &
{\rm for\ (pole/const.)}.
\label{c22p}
\end{eqnarray}

{}From (\ref{pb9a}), (\ref{pb10a}) and (\ref{c11})
the ratio of the vector and pseudoscalar decay constants
$f_{D_s^*}/f_{D_s}$ is given by
\begin{equation}
{f_{D_s^*}\over f_{D_s}}=1.36 \cdot
{|V_{cb}\, F_0^{BD}(m_{D_s}^2)|\over |V_{cb}\, F_1^{BD}(m_{D_s^{*}}^2)|}
\cdot
\left[
{{\cal{B}}({\bar{B}}^0\rightarrow D^+ D_s^{-*})\over
{\cal{B}}({\bar{B}}^0\rightarrow D^+ D_s^{-})} \right]^{{1/2}}
\cdot \left({0.87\over 0.96}\right) ,
\label{c23p}
\end{equation}
which gives
\begin{eqnarray}
{f_{D_s^*}\over f_{D_s}}=1.56\pm 0.49& &{\rm for\ (pole/pole)},
\nonumber\\
{f_{D_s^*}\over f_{D_s}}=1.43\pm 0.45& &{\rm for\ (pole/const.)}.
\label{c24p}
\end{eqnarray}

The decay constant is changed according to the $q^2$ behaviour of the
form factor $F_0(q^2)$. However the amount of change is less than $10 \%$
as shown in (\ref{c22p}). From this we know that the decay constant is not
so much dependent on the behaviour of the form factor.
Also when we consider the uncertainty from non-factorizable effects,
the decay constant is changed within $10 \%$ discrepancy.
In table \ref{table4} we show the results of $f_{D_s^*}, f_{D_s}$ and  
$f_{D_s^*}/f_{D_s}$ for different non-factorizable contributions.

As discussed in \cite{YYKeum}, 
when we consider the penguin contributions with non-factorizable effects,
the value of the decay constant $f_{D_s^{*}}$ is increased by $8 \%$, however,
for $f_{D_s}$ it is increased by up to $19 \%$. So the ratio 
$f_{D_s^{*}}/f_{D_s}$ is  decreased by $9 \%$.
In table 4 we summarized the values of decay constant $f_{D_s^{*}}$,
$f_{D_s}$ and the ratio of $f_{D_s^{*}}/f_{D_s}$ from various sources.
Our result for $f_{D_s}$ agrees well with
other theoretical calculations and experimental results within errors.
For the ratio  $f_{D_s^{*}}/f_{D_s}$,
our results have a value greater than 1,
however, Browder et al. \cite{browder} has a value less than 1.
It seems that this ratio is more likely to be greater than 1 when we consider
that the decay constant of $\rho$ meson is 1.5 times greater than that of
$\pi$ meson.
The difference of the results by Cheng and Yang \cite{Cheng-Yang}
comes from the different method and using different Wilson
coefficients.
Their values come by comparing two non-leptonic decay modes,
for instance ${\cal B}(B \rightarrow D D_s (D^{(*)}D_s^{*}))/
{\cal B}(B \rightarrow D \pi)$.

\vs{7mm}
{\bf \large 5. Conclusion}

By including the penguin contributions and the non-factorizable
term contributions,
we calculated the weak decay constants
$f_{D_s}$ and $f_{D_s^{*}}$ from
$\bar{B}^{0} \ra D^{+}\ell^{-} \nu$ and
$\bar{B}^{0} \ra D^{+}\bar{D}_s^{(*)}$. 
In our analysis, we consider the QCD-penguin 
and Electroweak-penguin contributions in hadronic two body decays
within the NDR renormalization scheme at next-to-leading order calculation.
We also considered the effect of two
different $q^2$-dependence of the form factor for  $F_0^{BD}(q^2)$.
The value of $f_{D_s}$ is  changed by less than $10 \%$ for 
different form factors.

The penguin effects for $B \ra DD_s$ decay is quite sizable, and we obtained
$f_{D_s} = 233 \pm 49$ ${\rm MeV}$ for the monopole type of $F_0^{BD}$,
$f_{D_s} = 255 \pm 54$ ${\rm MeV}$ for the constant $F_0^{BD}$. 
When we considered the non-factorizable contributions,
we obtained $f_{D_s^{*}} = 346 \pm 82$ ${\rm MeV}$ 
for the $D_s^{*}$ meson. 
These values will be improved
vastly when the large accumulated data samples are available
at the Belle and BaBar experiments in near future.

%\pagebreak
\vspace*{1.0cm}

\noindent
{\em Acknowledgements} \\
\indent
We are grateful to A. N. Kamal for reading this manuscript carefully.
YYK. would like to thank M. Kobayashi for his hospitality 
and encouragement.
This work was supported in part by Non-Directed-Research-Fund,
Korea Research Foundation 1997,
in part by the Basic Science Research Institute Program,
Ministry of Education, Project No. BSRI-97-2414, Korea,
and in part by the Grant-in Aid for Scientific
from the Ministry of Education, Science and Culture, Japan.\\

\pagebreak
%%%%%%%%%%%%%%%%%%%%%%%%%%%%%%%%%%%%%%%%%%%%%%%%%%%%%%%%%%%%%%%%%%%%%%%%%
\begin{table}[b]
\vspace*{0.5cm}
\hspace*{-1.0cm}
%\begin{center}
\begin{tabular}{|c|c|c|c|c|}   \hline
        &${\cal{B}} ({\bar{B}}^0\rightarrow D^+ \rho^-)$
        &${\cal{B}} ({\bar{B}}^0\rightarrow D^+ K^{-*})$
        &${\cal{B}} ({\bar{B}}^0\rightarrow D^+ \pi^-)$
        &${\cal{B}} ({\bar{B}}^0\rightarrow D^+ K^-)$
\\
        &$\times 10^3$&$\times 10^4$&$\times 10^3$&$\times 10^4$
\\   \hline
(pole/pole)    &$9.01\pm 1.54$&$4.62\pm 0.79$
               &$3.58\pm 0.61$&$2.74\pm 0.47$ \\
(pole/const.)  &$9.01\pm 1.54$&$4.62\pm 0.79$
               &$3.57\pm 0.61$&$2.71\pm 0.46$ \\ \hline
Experiments    &$8.4\pm 1.6\pm 0.7$&---
               &$3.1\pm 0.4\pm 0.2$&--- \\
\hline
\end{tabular}
%\end{center}
\caption{The obtained values of the branching ratios with $a_1 = 1.02$
and experimental measurements.
\label{table1}}
\end{table}
%%%%%%%%%%%%%%%%%%%%%%%%%%%%%%%%%%%%%%%%%%%%%%%%%%%%%%%%%%%%%%%%%

\begin{table}[ht]
\vspace*{0.5cm}
%\hspace*{-1.2cm}
\begin{center}
\begin{tabular}{|c||c|c||}   \hline
Coefficients  & Real Part & Imaginary Part \\
\hline
$C_1^{eff}$ &  1.168 & 0.0 \\ \hline
$C_2^{eff}$ & -0.365 & 0.0 \\ \hline
$C_3^{eff}$ &   2.25 $\cdot 10^{-2}$ & 4.5 $\cdot 10^{-3}$ \\ \hline
$C_4^{eff}$ &  -4.58 $\cdot 10^{-2}$ & -1.36 $\cdot 10^{-2}$ \\ \hline
$C_5^{eff}$ &   1.33 $\cdot 10^{-2}$ & 4.5  $\cdot 10^{-3}$ \\ \hline
$C_6^{eff}$ &  -4.80 $\cdot 10^{-2}$ & -1.36 $\cdot 10^{-2}$ \\ \hline
$C_7^{eff}$ &   2.37  $\cdot 10^{-4}$ & -2.88 $\cdot 10^{-4}$ \\ \hline
$C_8^{eff}$ &  4.30 $\cdot 10^{-4}$ & 0.0 \\ \hline
$C_9^{eff}$ &  -1.11 $\cdot 10^{-2}$ & -2.88  $\cdot 10^{-4}$ \\ \hline
$C_{10}^{eff}$ &  3.75 $\cdot 10^{-3}$ & 0.0 \\ \hline
\hline
\end{tabular}
\end{center}
\caption{The values of the effective wilson coefficient $C_{i}^{eff}$ 
with the $\mu = m_b(m_b) = 4,3$ ${\rm GeV}$, $m_c(m_b) = 0.95$ ${\rm GeV}$
in the NDR scheme at NLO calculation. 
\label{table2}}
\end{table}
%%%%%%%%%%%%%%%%%%%%%%%%%%%%%%%%%%%%%%%%%%%%%%%%%%%%%%%%%%%%%%%%%%%%%%%%%%%%%%%%%%
%%%%%%%%%%%%%%%%%%%%%%%%%%%%%%%%%%%%%%%%%%%%%%%%%%%%%%%%%%%%%%%%%

\begin{table}[ht]
\vspace*{0.5cm}
%\hspace*{-1.2cm}
\begin{center}
\begin{tabular}{|c||c|c||c|c||c|c||}   \hline
 &\multicolumn{2}{p{12em}||}{$(N_c)_{LL}=2$,\hspace{2mm} $(N_c)_{LR}=2$} 
 &\multicolumn{2}{p{12em}||}{$(N_c)_{LL}=2$,\hspace{2mm} $(N_c)_{LR}=5$} 
 &\multicolumn{2}{p{12em}||}
{$(N_c)_{LL}=3$,\hspace{2mm} $(N_c)_{LR}=3$} \\ \hline
Coeffs.  & Real Part & Imag. Part
 & Real Part & Imag. Part
 & Real Part & Imag. Part
 \\ \hline
$a_1$ 
&  0.985 & 0.0
&  0.985 & 0.0 
&  1.046 & 0.0
\\ \hline
$a_2$ 
&  0.219 & 0.0
&  0.219 & 0.0
&  0.024 & 0.0 \\ \hline
$a_3$ 
&  -4.00$\cdot 10^{-4}$ & -2.30 $\cdot 10^{-3}$
&  -4.00$\cdot 10^{-4}$ & -2.30 $\cdot 10^{-3}$
&   7.23$\cdot 10^{-3}$ & -3.30 $\cdot 10^{-5}$
 \\ \hline
$a_4$   
& -3.46 $\cdot 10^{-2}$ & -1.14$\cdot 10^{-2}$
& -3.46 $\cdot 10^{-2}$ & -1.14$\cdot 10^{-2}$
& -3.83 $\cdot 10^{-2}$ & -1.12$\cdot 10^{-2}$
 \\ \hline
$a_5$ 
&   -1.07 $\cdot 10^{-2}$ & 2.3 $\cdot 10^{-3}$ 
&   3.70 $\cdot 10^{-3}$ & 1.78 $\cdot 10^{-3}$
&   -2.70 $\cdot 10^{-3}$ & -3.33 $\cdot 10^{-5}$
\\ \hline
$a_6$ 
&  -4.13 $\cdot 10^{-2}$ & -1.14 $\cdot 10^{-2}$ 
&  -4.53 $\cdot 10^{-2}$ & -1.27 $\cdot 10^{-2}$ 
&  -4.36 $\cdot 10^{-2}$ & -1.21 $\cdot 10^{-2}$ 
\\ \hline
$a_7$ 
&  -2.19 $\cdot 10^{-5}$ & -2.88 $\cdot 10^{-4}$ 
&  -1.51 $\cdot 10^{-4}$ & -2.88 $\cdot 10^{-4}$ 
&  -9.35 $\cdot 10^{-5}$ & -2.88 $\cdot 10^{-4}$ 
\\ \hline
$a_8$ 
&   3.11 $\cdot 10^{-4}$ & -1.44 $\cdot 10^{-4}$ 
&  -3.82 $\cdot 10^{-4}$ & -5.77 $\cdot 10^{-5}$ 
&   3.51 $\cdot 10^{-4}$ & -9.61 $\cdot 10^{-5}$ 
\\ \hline
$a_9$ 
&  -9.27 $\cdot 10^{-3}$ & -2.88 $\cdot 10^{-4}$ 
&  -9.27 $\cdot 10^{-3}$ & -2.88 $\cdot 10^{-4}$ 
&  -9.90 $\cdot 10^{-3}$ & -2.88 $\cdot 10^{-4}$ 
\\ \hline
$a_{10}$ 
&  -1.82 $\cdot 10^{-4}$ & -1.44 $\cdot 10^{-4}$
&  -1.82 $\cdot 10^{-4}$ & -1.44 $\cdot 10^{-4}$ 
&   3.39 $\cdot 10^{-5}$ & -9.61 $\cdot 10^{-5}$
\\ \hline
\hline
\end{tabular}
\end{center}
\caption{The values of the effective coefficients $a_i$
with $\mu = m_b(m_b) = 4,3$ ${\rm GeV}$ and $m_c(m_b) = 0.95$ ${\rm GeV}$
in the NDR scheme at NLO calculation.
$a_{2i}$ and
$a_{2i-1}$ are defined by
$a_{2i-1} = C_{2i-1}^{eff} + C_{2i}^{eff}/N_c^{eff}$ and
$a_{2i} = C_{2i}^{eff} + C_{2i-1}^{eff}/N_c^{eff}$. Here we have taken
$(N_c)_{LL}$ for (V-A)(V-A) interaction and $(N_c)_{LR}$ for
(V-A)(V+A) interaction. 
\label{table3}}
\end{table}
%%%%%%%%%%%%%%%%%%%%%%%%%%%%%%%%%%%%%%%%%%%%%%%%%%%%%%%%%%%%%%%%%%%%%%%%%%%%%%%%%%
\begin{table}[ht]
\vspace*{0.5cm}
%\hspace*{-1.2cm}
\begin{center}
\begin{tabular}{|c|c|c||c|c|c|}   \hline
   &$(N_c)_{LL}$ & $(N_c)_{LR}$ &$f_{D_s^*}$ (${\rm MeV}$)
   &$f_{D_s}$ (${\rm MeV}$)&$f_{D_s^*}/f_{D_s}$ \\
\hline \hline
   &  2 & 2 &$346\pm 82$&$231\pm 48$&$1.57\pm 0.50$ \\
\cline{2-6}
(pole/pole) &2 & 5 & $346\pm 82$&$233\pm 49$&$1.56\pm 0.49$ \\
\cline{2-6}
 &3 & 3 & $325\pm 77$&$216\pm 45$&$1.57\pm 0.50$ \\
\hline \hline
   &  2 & 2 &$346\pm 82$&$252\pm 53$&$1.44\pm 0.45$ \\
\cline{2-6}
(pole/const.) &2 & 5 & $346\pm 82$&$255\pm 54$&$1.43\pm 0.45$ \\
\cline{2-6}
 &3 & 3 & $325\pm 77$&$235\pm 50$&$1.44\pm 0.45$ \\
\hline \hline
\multicolumn{3}{|c||}{ Browder $et \,\, al.$\cite{browder}}
&$243\pm 70$&$277\pm 77$ &$0.88 \pm 0.35$ \\ 
\multicolumn{3}{|c||}{Hwang and Kim  \cite{hk1}}
&$362\pm 15$&$309\pm 15$&$1.17\pm 0.02$ \\
\multicolumn{3}{|c||}{Cheng and Yang  \cite{Cheng-Yang}}
&$266\pm 62$&$261\pm 46$&$1.02\pm 0.30$ \\
\hline
\multicolumn{3}{|c||}
{Capstick and Godfrey \cite{cg}}
& &$290\pm 20$& \\
\multicolumn{3}{|c||}
{Dominguez \cite{doming}}
& &$222\pm 48$& \\
\multicolumn{3}{|c||}
{UKQCD \cite{UKQCD}}
& &$212^{+4+46}_{-3-7}$& \\
\multicolumn{3}{|c||}
{BLS \cite{BLS}}
& &$230\pm 7\pm 35$&\\
\multicolumn{3}{|c||}
{MILC \cite{MILC}}
&&$199\pm 8^{+40+10}_{-11-0}$&\\
\multicolumn{3}{|c||}
{Becirevic $et \,\, al.$ \cite{Bec}}
&$272\pm 16^{+0}_{-20}$& $231\pm 12^{+6}_{-0}$&$1.18 \pm 0.18$ \\
\hline
\multicolumn{3}{|c||}
{WA75 \cite{wa75}}&
&$238 \pm 47 \pm 21 \pm 48 $& \\
\multicolumn{3}{|c||}
{CLEO 1 \cite{cleo94}} &
&$282 \pm 30 \pm 43 \pm 34 $& \\
\multicolumn{3}{|c||}
{CLEO 2 \cite{cleo95}} &
&$280 \pm 19 \pm 28 \pm 34 $& \\
\multicolumn{3}{|c||}
{BES \cite{bes}} &
&$430^{+ 150}_{-130} \pm 40 $& \\
\multicolumn{3}{|c||}
{E653 \cite{E653}} &
&$190 \pm 34 \pm 20 \pm 26 $& \\
\hline
\end{tabular}
\end{center}
\caption{The obtained values of
$f_{D_s^*}$ (${\rm MeV}$) and $f_{D_s}$ (${\rm MeV}$), and their ratio
$f_{D_s^*}/f_{D_s}$, and the results from other theoretical
calculations and existing experimental results.
Here we refered the corrected $f_{D_s}$ values \cite{cleo98}
for the experimental data \cite{wa75} - \cite{E653}.
\label{table4}}
\end{table}
%%%%%%%%%%%%%%%%%%%%%%%%%%%%%%%%%%%%%%%%%%%%%%%%%%%%%%%%%%%%%%%%%%


\vspace{20mm}
\begin{thebibliography}{99}

\bibitem{rosner} J. L. Rosner, in {\bf Particles and Fields 3},
Proceedings of the 1988 Banff Summer Inst.,
Banff, Alberta,Canada, ed. by A.N. Kamal and F.C. Khanna,
World Scientific. Singapore, 1989, 395.

\bibitem{bort-stone}
D. Bortoletto and S. Stone, Phys. Rev. Lett. {\bf 65}, 2951 (1990).

\bibitem{HF}
M. Neubert, V. Rieckert, B. Stech and Q. P. Xu, in {\bf Heavy
Flavours} ed. by A. J. Buras and M. Lindner,
World Scientific. Singapore, 1992

\bibitem{YYKeum}
Y.-Y. Keum, hep-ph/9810369

\bibitem{GKKP}
M. Gourdin, A.N. Kamal, Y.-Y. Keum and X.Y. Pham,
Phys. Lett B {\bf 333}, 507 (1994).

\bibitem{Hai-Yang} H.Y. Cheng, H.N. Li, and K.C. Yang,
hep-ph/9902239.

\bibitem{wsb} M. Wirbel, B. Stech and M. Bauer,
Z. Phys. C {\bf 29}, 637 (1985);
M. Bauer and M. Wirbel, Z. Phys. C {\bf 42}, 671 (1989).

\bibitem{ks} J.G. Korner and G.A. Schuler,
Z. Phys. C {\bf 38}, 511 (1988);
$ibid.$ (erratum) C {\bf 41}, 690 (1989).

\bibitem{iw} N. Isgur and M.B. Wise, Phys. Lett. B {\bf 232}, 113 (1989);
$ibid.$ B {\bf 237}, 527 (1990).

\bibitem{nr} M. Neubert and V. Rieckert,
Nucl. Phys. B {\bf 382}, 97 (1992).

\bibitem{rpp} Review of Particle Physics, R.M. Barnett et al.,
Phys. Rev. D {\bf 54}, 1 (1996).

\bibitem{stone} S. Stone, hep-ph/9610305,
in Proceedings of NATO Advanced Study Institute on Techniques and
Concepts of High Energy Physics, Virgin Islands, July 1996.

\bibitem{burasNDR} A. J. Buras, Nucl. Phys. B {\bf 434}, 606 (1995).

\bibitem{deshpande} N.G. Deshpande and X.-G. He,
Phys. Rev. Lett. {\bf 74}, 26 (1995).

\bibitem{fleischer} R. Fleischer, Z. Phys. C {\bf 58}, 483 (1993);
C {\bf 62}, 81 (1994).

\bibitem{BS} A.J. Buras and L. Silverstini, 
hep-ph/9806278.

\bibitem{browder} T.E. Browder, K. Honscheid and D. Pedrini,
hep-ph/9606354, UH-515-848-96, OHSTPY-HEP-E-96-006,
To appear in
{\it Annual Review of Nuclear and Particle Science, Vol. 46}.

\bibitem{hk1} D.S. Hwang and G.-H. Kim,
Phys. Rev. D {\bf 55}, 6944 (1997).

\bibitem{Cheng-Yang} H.Y. Cheng and K.C. Yang,
hep-ph/9811249.

\bibitem{cg} S. Capstick and S. Godfrey,
Phys. Rev. D {\bf 41}, 2856 (1990).

\bibitem{doming} C.A. Dominguez, in Proceedings of the Third
Workshop on Tau-Charm Factory, Marbella, Spain, June, 1993.

\bibitem{UKQCD} R.M. Boxter et al., UKQCD Collaboration,
Phys. Rev. D {\bf 49}, 1594 (1994).

\bibitem{BLS} C.W. Bernard, J.N. Labrenz and A. Soni,
Phys. Rev. D {\bf 49}, 2536 (1994).

\bibitem{MILC} C. Bernard et al., MILC Collaboration,
hep-ph/9709328.

\bibitem{Bec} D. Becirevic, Ph. Boucaud, J.P. Leroy, V. Lubicz,
G. Martinelli, F. Mescia and F. Rapuano,
hep-lat/9811003.

\bibitem{wa75} S. Aoki et al., WA75 Collaboration,
Prog. Theor. Phys. {\bf 89}, 131 (1993).

\bibitem{cleo94} D. Acosta et al., CLEO Collaboration, Phys. Rev.
D {\bf 49}, 5690 (1994).

\bibitem{cleo95} D. Gibaut et al., CLEO Collaboration,
CLEO CONF 95-22, EPS0184 (1995).

\bibitem{bes} J.Z. Bai et al., BES Collaboration,
Phys. Rev. Lett. {\bf 74}, 4599 (1995).

\bibitem{E653} K. Kodama et al., E653 Collaboration,
Phys. Lett. B {\bf 382}, 299 (1996).

\bibitem{cleo98} M. Chadha et al., CLEO Collaboration,
Phys. Rev. D {\bf 58}, 32002 (1998).

\end{thebibliography}
\end{document}